# Tracking Group Evolution in Social Networks

Piotr Bródka, Stanisław Saganowski, Przemysław Kazienko

[1] Wrocław University of Technology, Wyb.Wyspiańskiego 27, 50-370 Wrocław, Poland
[2] Research Engineering Center Sp. z o.o., ul. Strzegomska 46B, 53-611 Wrocław, Poland
`piotr.brodka@pwr.wroc.pl,`
`stanislaw.saganowski@student.pwr.wroc.pl, kazienko@pwr.wroc.pl`

**Abstract.** Easy access and vast amount of data, especially from long period of time, allows to divide social network into timeframes and create temporal social network. Such network enables to analyse its dynamics. One aspect of the dynamics is analysis of social communities evolution, i.e., how particular group changes over time. To do so, the complete group evolution history is needed. That is why in this paper the new method for group evolution extraction called *GED* is presented.

## 1  Introduction

One of the areas of science which in recent years is rapidly growing is social network analysis. One of the main reasons for this is growing number of different social networking systems and growth of the Internet together with simple and continuous way to obtain data from which we can extract those social networks.

Group extraction and their evolution are among the topics which arouse the greatest interest in the domain of social network analysis. However, while the grouping methods in social networks are developed very dynamically, the methods of group evolution discovery and analysis are still '*uncharted territory*' on the social network analysis map. In recent years only few methods for tracking changes of social groups have been proposed: [2], [3]. [5], [6]. Therefore in this paper the new method for the group evolution discovery called *GED* is proposed, analysed and compared with two methods by Asur and by Palla. It should also be mentioned that this article is an extension and continuation of research presented in [1].

## 2  Group Evolution

Before the method can be presented, it is necessary to describe a few concepts related to social networks:

Temporal social network *TSN* a list of succeeding timeframes (time windows) *T*. Each timeframe is in fact one social network $SN(V,E)$ where $V$ – is a set of vertices and $E$ is a set of directed edges $\langle x,y\rangle : x,y \in V, x \neq y$

$$TSN = <T_1, T_2, \ldots T_m>, \quad m \in N$$
$$T_i = SN_i(V_i, E_i), \quad i = 1, 2, \ldots, m \tag{1}$$
$$E_i = <x, y>: x, y \in V_i, x \neq y \quad i = 1, 2, \ldots, m$$

Evolution of particular social community can be represented as a sequence of events (changes) following each other in the successive time windows (timeframes) within the temporal social network. Possible events in social group evolution are:

1. *Continuing* (stagnation) – the group continue its existence when two groups in the consecutive time windows are identical or when two groups differ only by few nodes but their size remains the same.
2. *Shrinking* – the group shrinks when some nodes has left the group, making its size smaller than in the previous time window. Group can shrink slightly i.e. by a few nodes or greatly losing most of its members.
3. *Growing* (opposite to shrinking) – the group grows when some new nodes have joined the group, making its size bigger than in the previous time window. A group can grow slightly as well as significantly, doubling or even tripling its size.
4. *Splitting* – the group splits into two or more groups in the next time window when few groups from timeframe $T_{i+1}$ consist of members of one group from timeframe $T_i$. We can distinguish two types of splitting: (1) *equal,* which means the contribution of the groups in split group is almost the same and (2) *unequal* when one of the groups has much greater contribution in the split group, which for this one group the event might be similar to shrinking.
5. *Merging*, (reverse to splitting) – the group has been created by merging several other groups when one group from timeframe $T_{i+1}$ consist of two or more groups from the previous timeframe $T_i$. Merge, just like the split, might be (1) *equal*, which means the contribution of the groups in merged group is almost the same, or (2) unequal, when one of the groups has much greater contribution into the merged group. In second case for the biggest group the merging might be similar to growing.
6. *Dissolving* happens when a group ends its life and does not occur in the next time window, i.e., its members have vanished or stop communicating with each other and scattered among the rest of the groups.
7. *Forming* (opposed to dissolving) of new group occurs when group which has not existed in the previous time window $T_i$ appears in next time window $T_{i+1}$. In some cases, a group can be inactive over several timeframes, such case is treated as dissolving of the first group and forming again of the, second, new one.

## 3    Tracking Group Evolution in Social Networks

The *GED* method, to match two groups from consecutive timeframes takes into consideration both, the quantity and quality of the group members. To express group members quality one of the centrality measures may be used. In this article authors have decided to utilize social position (*SP*) measure [4] to reflect the quality of group members.

To track social community evolution in social network the new method called *GED* (Group Evolution Discovery) was developed. Key element of this method is a new measure called *inclusion*. This measure allows to evaluate the inclusion of one group in another. The inclusion of group $G_1$ in group $G_2$ is calculated as follows:

$$I(G_1, G_2) = \overbrace{\frac{|G_1 \cap G_2|}{|G_1|}}^{group\,quantity} \cdot \underbrace{\frac{\sum_{x \in (G_1 \cap G_2)} SP_{G_1}(x)}{\sum_{x \in (G_1)} SP_{G_1}(x)}}_{group\,quality} \qquad (2)$$

Naturally, instead of social position (*SP*) any other measure which indicates user importance can be used e.g. centrality degree, betweenness degree, page rank etc. But it is important that this measure is calculated for the group and not for social network in order to reflect node position in group and not in the whole social network.

As mentioned earlier the *GED* method, used to track group evolution, takes into account both the quantity and quality of the group members.

The quantity is reflected by the first part of the *inclusion* measure, i.e. what portion of $G_1$ members is shared by both groups $G_1$ and $G_2$, whereas the quality is expressed by the second part of the *inclusion* measure, namely what contribution of important members of $G_1$ is shared by both groups $G_1$ and $G_2$. It provides a balance between the groups, which contain many of the less important members and groups with only few but key members.

It is assumed that only one event may occur between two groups ($G_1$, $G_2$) in the consecutive timeframes, however one group in timeframe $T_i$ may have several events with different groups in $T_{i+1}$.

---

**GED – Group Evolution Discovery Method**

**Input:** *TSN* in which at each timeframe $T_i$ groups are extracted by any community detection algorithm. Calculated any user importance measure.

1. For each pair of groups <$G_1$, $G_2$> in consecutive timeframes $T_i$ and $T_{i+1}$ inclusion of $G_1$ in $G_2$ and $G_2$ in $G_1$ is counted according to equations (3).

2. Based on inclusion and size of two groups one type of event may be assigned:

    a. *Continuing*: $I(G_1,G_2) \geq \alpha$ and $I(G_2,G_1) \geq \beta$ and $|G_1| = |G_2|$

    b. *Shrinking*: $I(G_1,G_2) \geq \alpha$ and $I(G_2,G_1) \geq \beta$ and $|G_1| > |G_2|$ OR $I(G_1,G_2) < \alpha$ and $I(G_2,G_1) \geq \beta$ and $|G_1| \geq |G_2|$ OR $I(G_1,G_2) \geq \alpha$ and $I(G_2,G_1) < \beta$ and $|G_1| \geq |G_2|$ and there is only one match between $G_1$ and groups in the next time window $T_{i+1}$

    c. *Growing*: $I(G_1,G_2) \geq \alpha$ and $I(G_2,G_1) \geq \beta$ and $|G_1|<|G_2|$ OR $I(G_1,G_2) \geq \alpha$ and $I(G_2,G_1) < \beta$ and $|G_1| \leq |G_2|$ OR $I(G_1,G_2) < \alpha$ and $I(G_2,G_1) \geq \beta$ and $|G_1| \leq |G_2|$ and there is only one match between $G_2$ and groups in the next previous window $T_i$

    d. *Splitting*: $I(G_1,G_2) < \alpha$ and $I(G_2,G_1) \geq \beta$ and $|G_1| \geq |G_2|$ OR $I(G_1,G_2) \geq \alpha$ and

> $I(G_2,G_1) < \beta$ and $|G_1| \geq |G_2|$ and there is more than one match between $G_1$ and groups in the next time window $T_{i+1}$
>
> e. *Merging*: $I(G_1,G_2) \geq \alpha$ and $I(G_2,G_1) < \beta$ and $|G_1| \leq |G_2|$ OR $I(G_1,G_2) < \alpha$ and $I(G_2,G_1) \geq \beta$ and $|G_1| \leq |G_2|$ and there is more than one match between $G_2$ and groups in the previous time window $T_i$
>
> f. *Dissolving*: for $G_1$ in $T_i$ and each group $G_2$ in $T_{i+1}$   $I(G_1,G_2) < 10\%$ and $I(G_2,G_1) < 10\%$
>
> g. *Forming*: for $G_2$ in $T_{i+1}$ and each group $G_1$ in $T_i$   $I(G_1,G_2) < 10\%$ and $I(G_2,G_1) < 10\%$

The scheme which facilitate understanding of the event selection for the pair of groups in the method is presented in Figure 1.

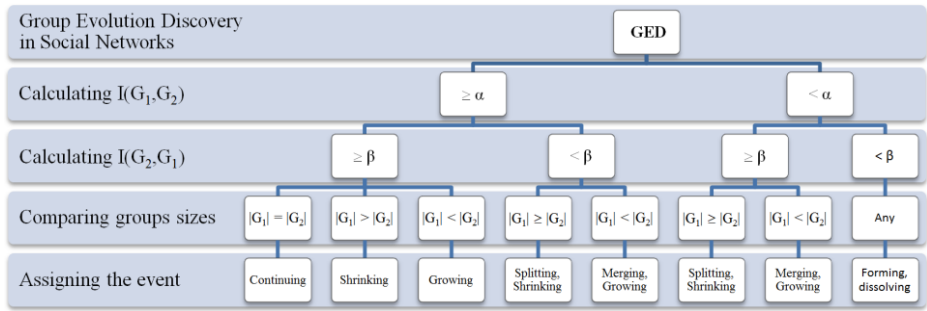

**Fig. 1.**.The decision tree for assigning the event type to the group

The indicators *α* and *β* are the *GED* method parameters which can be used to adjust the method to particular social network and community detection method. After the experiments analysis authors suggest that the values of *α* and *β* should be from range [50%;100%]

**Acknowledgments.** The work was supported by: Fellowship co-Financed by the European Union within the European Social Fund, The Polish Ministry of Science and Higher Education, the research project, 2010-13, The training in "Green Transfer" co-financed by the EU from the European Social Fund

**Extended version of this paper:** Bródka P., Saganowski P., Kazienko P.: *GED: The Method for Group Evolution Discovery in Social Networks*, Social Network Analysis and Mining, DOI:10.1007/s13278-012-0058-8